 \documentstyle[11pt,aaspp4]{article} % preprint   (single-space) style
 \slugcomment{Accepted for publication in 
             {\it Ap.J.\/} 1999 February 20}
\lefthead{Mauche}
\righthead{ORFEUS and IUE Spectra of EX Hya}

\newcommand\Mwd  {M_{\rm wd}}
\newcommand\Rwd  {R_{\rm wd}}
\newcommand\Msun {{\rm M_{\odot}}}
\newcommand\lax{{\lower0.75ex\hbox{$<$}\atop\raise0.5ex\hbox{$\sim$}}}
\newcommand\gax{{\lower0.75ex\hbox{$>$}\atop\raise0.5ex\hbox{$\sim$}}}

\begin{document}

\title{ORFEUS II and IUE Spectroscopy of EX Hydrae}

\author{Christopher W.\ Mauche\altaffilmark{1}}
\affil{Lawrence Livermore National Laboratory, \\
       L-43, 7000 East Avenue, Livermore, CA 94550; \\
       mauche@cygnus.llnl.gov}

\altaffiltext{1}{{\it ORFEUS-SPAS II\/} Guest Investigator}

\clearpage % force page break; deleted when you change from pp to ms

% Abstract
%---------------------------------------------------------

\begin{abstract}

Using five high-resolution ($\lambda/\Delta\lambda\approx 3000$) FUV
($\lambda\lambda=910$--1210~\AA ) spectra acquired in 1996 December 
during the flight of the {\it ORFEUS-SPAS II\/} mission, 47 archival {\it
IUE\/} short-wavelength UV ($\lambda\lambda=1150$--1950~\AA ) spectra, 
and archival {\it EUVE\/} deep survey photometry ($\lambda\lambda\approx
70$--200~\AA ), we present a detailed picture of the behavior of the
magnetic cataclysmic variable EX Hydrae in the vacuum ultraviolet. Like
{\it HUT\/} spectra of this source, these FUV and UV spectra reveal broad
emission lines of He II, C~II--IV, N~III and V, O~VI, Si~III--IV, and
Al~III superposed on a continuum which is blue in the UV and nearly
flat in the FUV. Like {\it ORFEUS\/} spectra of AM~Her, the O~VI
$\lambda\lambda 1032$, 1038 doublet is resolved into broad ($\rm
FWHM\approx 2000~km~s^{-1}$) and narrow ($\rm FWHM\approx
200~km~s^{-1}$) emission components. Consistent with its behavior in the
optical, and hence consistent with the accretion curtain model of this
source, the FUV and UV continuum flux densities, the FUV and UV broad
emission line fluxes, and the radial velocity of the O~VI broad emission
component all vary on the spin phase of the white dwarf, with the maximum
of the FUV and UV continuum and broad emission line flux light curves
coincident with maximum {\it blueshift\/} of the broad O~VI emission
component. On the binary phase, the strong eclipse of the EUV flux by
the bulge on the edge of the accretion disk is accompanied by narrow and
relatively weak absorption components of the FUV emission lines and
30\%--40\% eclipses of all the UV emission lines except He~II $\lambda
1640$, while the UV continuum is largely unaffected. Furthermore, both
the flux and radial velocity of the O~VI narrow emission component vary
with binary phase. The relative phasing of the FUV and UV continuum light
curves and the FUV emission-line radial velocities implicate the
accretion funnel as the source of the FUV and UV continuum and the O~VI
broad emission component, and the white dwarf as the source of the O~VI
narrow emission component. Various lines of evidence imply that the
density of both the broad- and narrow-line regions is $n\gax10^{11}~\rm
cm^{-3}$, but the O~VI line ratios imply that the narrow-line region is
optically thick while the broad-line region is more likely optically thin.

\end{abstract}

\keywords{binaries: close ---
          stars: individual (EX Hydrae) ---
          stars: magnetic fields ---
          ultraviolet: stars ---
          white dwarfs}

\clearpage % force page break; delete when you change from pp to ms
 
% Body of the paper
%---------------------------------------------------------

\section{Introduction}

Among the classes of cataclysmic variables (CVs), there is one in which
the magnetic field of the white dwarf is strong enough to influence the
flow of material lost by the secondary. These magnetic CVs are subdivided 
into the spin-synchronous ($P_{\rm spin}\approx P_{\rm orb}$) polars 
(AM~Her stars) and the spin-asynchronous ($P_{\rm spin}< P_{\rm orb}$)
intermediate polars (DQ~Her stars). In polars, the accreting matter is
channeled along the magnetic field lines for most of its trajectory from
the secondary's inner Lagrange point to the white dwarf surface, while in
intermediate polars (IPs), accretion is moderated by a disk. Although the
disk maintains an essentially Keplerian velocity profile at large radii,
it is truncated at small radii where the magnetic stresses become large
enough to remove the disk angular momentum in a radial distance which is
small compared to the distance to the white dwarf; the material then
leaves the disk and follows the magnetic field lines down to the white
dwarf surface in a manner similar to that of polars. In either class of
magnetic CVs, the velocity of the flow as it nears the white dwarf is
highly supersonic [$v = 3600\, (\Mwd/0.5\,\Msun )^{1/2} (\Rwd/10^9~{\rm
cm})^{-1/2}~\rm km~s^{-1}$], hence to match boundary conditions, the flow
must pass through a strong shock far enough above the white dwarf surface
for the hot [$kT = 16\, (\Mwd/0.5\,\Msun )(\Rwd/10^9~{\rm cm})^{-1}$
keV], post-shock flow to be decelerated by pressure forces and settle on
the white dwarf surface. Magnetic CVs are therefore strong X-ray sources
modulated at the spin period of the white dwarf. For additional details
of magnetic CVs, see Cropper (1990), Patterson (1994), and Warner (1995);
for recent results, see the volumes by Buckley \& Warner (1995) and
Hellier \& Mukai (1999).

EX~Hydrae is a bright ($V\approx 13$), high-inclination ($i=77^\circ\pm
1^\circ$), eclipsing IP with an orbital period of 98.26 minutes and a
white dwarf spin period of 67.03 minutes. The mass of the white dwarf
is measured by both dynamical (\cite{hel96}) and X-ray spectroscopic
(\cite{fuj97}) methods to be $M_{\rm wd}=0.49\, \Msun $, while details
of the accretion geometry are established by the optical and X-ray
observations of Hellier et~al.\ (1987) and Rosen, Mason, \& C\'ordova
(1988). In the resulting ``accretion curtain'' model of EX~Hya
specifically and IPs in general, the spin-phase modulations are the
result of the angular offset between the spin and magnetic dipole pole
axes and the consequent strong azimuthal asymmetry of the flow of
material from the disk to the surface of the white dwarf. Because of
absorption by this accretion curtain, the spin-phase light curves peak
when the upper pole points {\it away\/} from the observer---when the
blueshift of the emission lines is maximum. In addition to this 
spin-phase modulation, binary-phase modulations are produced by partial
eclipses by the secondary and by the bulge on the edge of the accretion
disk.

EX~Hya has been studied extensively in the X-ray and optical wavebands
(in addition to the references above, see, e.g., \cite{sie89};
\cite{ros91}; \cite{all98}; \cite{muk98}), but less so at UV and FUV
wavelengths. Despite the 174 {\it International Ultraviolet Explorer\/}
({\it IUE\/}) spectra of EX~Hya in the archive, the discussion of these
UV data has been limited to the papers by Bath, Pringle, \& Whelan (1980,
written before EX~Hya was recognized as an IP), Krautter \& Buchholz
(1990, a 2-page paper in a conference proceedings), and the statistical
studies of Verbunt (1987), la Dous (1991), and Mauche, Lee, \& Kallman
(1997). Greeley et~al.\ (1997) recently described and modeled UV--FUV
spectra of this source acquired with the {\it Hopkins Ultraviolet
Telescope\/} ({\it HUT\/}) during the Astro-2 mission in 1995 March. To
extend the effort of documenting the phenomenology and understanding the
accretion geometry of EX~Hya, we here analyze and discuss five FUV 
spectra of this source acquired in 1996 December during the flight of 
the {\it Orbiting and Retrievable Far and Extreme Ultraviolet
Spectrograph--Shuttle Pallet Satellite\/} ({\it ORFEUS-SPAS\/}) {\it
II\/} mission. These spectra are superior those of {\it HUT\/} because 
of the higher spectral resolution and more extensive phase coverage, but
suffer from the narrower bandpass. To help offset this shortcoming, we
make use of an extensive set of {\it IUE\/} spectra of EX~Hya obtained 
in 1995 June by K.~Mukai. For completeness and ease of reference, we 
also present and describe the EUV/soft X-ray spin- and binary-phase 
light curves of EX~Hya measured by {\it EUVE\/} in 1994 May--June
(\cite{hur97}). AAVSO measurements during and near the times of these
observations demonstrate that in each instance EX~Hya was at its
quiescent optical magnitude of $V\approx 13$ (\cite{mat98}).

\section{EUVE Photometry}

As described in detail by Hurwitz et~al.\ (1997), EX~Hya was observed by
{\it EUVE\/} for 150 kiloseconds beginning on 1994 May 26.\footnote{For
the record, note that Hurwitz et~al.\ (1997) erroneously report that the
{\it EUVE\/} observation of EX~Hya began on 1994 May 29; this date is
actually the midpoint of the observation. Similarly, the dates referred
to in \S 2.1 (\S 2.3) of that paper are not the first, but the $\approx
48$th ($\approx 70$th) observed binary eclipse (spin maximum).} The
resulting deep survey ($\lambda\lambda\approx 70$--200~\AA ) spin- and
binary-phase count rate light curves are shown in Figure~1, where the 
ephemerides of Hellier \& Sproats (1992) have been employed to convert HJD
photon arrival times to white dwarf spin and binary phases.\footnote{The
sinusoidal term in the binary ephemeris of Hellier \& Sproats (1992) is
ignored here and elsewhere in this paper because it is uncertain and
because it has a full range of only 48 seconds or 0.008 binary cycles;
at the midpoint of the {\it EUVE\/} observation, the correction amounts
to $-41$ seconds or $-0.007$ binary cycles.} Due to the $\sim 30$\%
efficiency of {\it EUVE\/} observations, the photons from whence these
light curves were constructed were acquired over an interval of 6.5
days (from HJD 2449498.89420 until 2449505.42824). This observation was
therefore sufficiently long to avoid the spin- and binary-period aliasing
typical of low-Earth-orbit satellite observations of EX~Hya, including
the {\it ORFEUS\/} and {\it IUE\/} observations described below.

The binary-phase {\it EUVE\/} light curve of EX~Hya is shown in the
middle and lower panels of Figure~1 and is seen to manifest a broad dip
at $\phi _{98}\approx 0.85^{+0.15}_{-0.25}$ and a narrow eclipse at 
$\phi _{98}\approx 0.97$. The dip is understood to be due to the passage 
through the line of sight of the bulge on the edge of the accretion disk
caused  by the impact of the accretion stream. With a residual intensity
of approximately 0.13, the optical depth of the bulge is $\tau\approx 
2.0$ at $\lambda\approx 90$~\AA , the peak of the effective area curve of 
the deep survey instrument. If the occulting material is neutral and 
has solar abundances (specifically, one He atom for every ten H atoms), 
the inferred column density is $N_{\rm H}\approx 7.4\times 10^{19}~\rm
cm^{-2}$. Such a column is essentially transparent ($\tau\le 0.01$) above
1.2 keV, consistent with the fact that the dip is seen only in soft
X-rays (e.g., \cite{ros88}). The narrow eclipse is understood to be due 
to the grazing occultation of the EUV/soft X-ray emission region by the
secondary. Fitting a linear background minus a Gaussian to the interval
$\phi _{98}=0.95$--0.99, we find that the eclipse is centered at $\phi
_{98} = 0.9714\pm 0.0003$, has a $\rm FWHM=0.007\pm 0.001$ or $38\pm 6$
seconds, and a full width of $\Delta\phi _{98}\sim 0.01$ or 60 seconds;
the residual intensity at mid-eclipse is consistent with zero at $0.014
\pm 0.013~\rm counts~s^{-1}$. In contrast, the eclipse by the secondary
of the hard X-ray emission region is significantly wider ($\Delta
\phi_{98} \approx 0.03$ or 180 seconds) and partial (eclipse depth =
30--60\%; \cite{beu88}; \cite{ros91}; \cite{muk98}). The centroid of 
the hard X-ray eclipse was recently measured with {\it RXTE\/} by Mukai
et~al.\ (1998) to be centered at $\phi _{98}= 0.98$, reinforcing the
impression that the binary ephemeris of Hellier \& Sproats (1992) may
need to be updated.

The spin-phase {\it EUVE\/} light curve of EX~Hya is shown in the upper
panel of Figure~1 and is seen to vary more sharply than a sine wave, but
it is nonetheless reasonably well approximated by the sinusoidal function
$A +B\, \sin 2\pi (\phi _{67}-\phi _0)$ with $A=0.158 \pm 0.001~\rm
counts~s^{-1}$, $B = 0.105\pm 0.002~\rm counts~s^{-1}$, and $\phi _0 =
0.790\pm 0.002$. The relative pulse amplitude is therefore $B/A = 67\% 
\pm 1\%$ and the light curve peaks at $\phi _{67} = 0.040\pm 0.002$. This
phasing again is formally different from the ephemeris of Hellier \&
Sproats (1992), but it establishes to sufficient accuracy for the present
purposes that the EUV/soft X-ray light curve peaks at $\phi _{67}\approx
0$.

\section{ORFEUS Spectroscopy}

The FUV spectra of EX~Hya were acquired with the Berkeley spectrograph
in the {\it ORFEUS\/} telescope during the flight of the {\it ORFEUS-SPAS
II\/} mission in 1996 November--December. The general design of the 
spectrograph is discussed by Hurwitz \& Bowyer (1986, 1996), while
calibration and performance of the {\it ORFEUS-SPAS II\/} mission are
described by Hurwitz et~al.\ (1998); for the present purposes, it is
sufficient to note that the spectra cover the range $\lambda\lambda =
910$--1210~\AA \ and that the mean instrument profile $\rm FWHM\approx
0.33$~\AA , hence $\lambda/\Delta\lambda\approx 3000$. Acquisition of the
{\it ORFEUS\/} exposures was complicated by the fact that the satellite
period (91 min) nearly equals the binary orbital period and four thirds
of the white dwarf spin period. After consulting with B.~Greeley it was
decided to concentrate on the spin period, with observations every
satellite orbit for 6 orbits, but practical considerations resulted in
the coverage shown in Figure~1 and detailed in Table~1, which lists the
HJD of the start of the exposures, the length of the exposures, and the
range of binary and spin phases assuming the ephemerides of Hellier \&
Sproats (1992).

Figure~2 shows the background-subtracted and flux-calibrated {\it
ORFEUS\/} spectra binned to a resolution of 0.1~\AA \ and smoothed with a
5-point triangular filter. Relatively strong residual geocoronal emission
lines of H~I $\lambda 1025.7$ (Lyman $\beta $), He~I $\lambda 584.3$ (at
1168.7~\AA \ in second order), N~I $\lambda 1134$, $\lambda 1200$, and
O~I $\lambda 988.7$ have been subtracted from these spectra by fitting
Gaussians in the neighborhood ($\pm 5$~\AA ) of each line. The remaining
geocoronal emission lines are all very weak and contaminate only a
limited number of discrete ($\rm FWHM\approx 0.8$~\AA ) portions of the
spectra. These FUV spectra are generally consistent with the {\it HUT\/}
spectra acquired 1995 March (\cite{gre97}), with emission lines of O~VI
$\lambda\lambda 1032$, 1038 and C~III $\lambda 977$, $\lambda 1176$
superposed on a nearly flat continuum. The broad and variable emission
feature at $\lambda\approx 990$~\AA \ is likely N~III, but the flux and
position of this feature are uncertain because it coincides with a strong
increase in the background at $\lambda\approx 1000$~\AA \ which renders
noisy the short-wavelength end of these spectra.

To quantify the continuum flux density variations of the FUV spectra of
EX~Hya, we measured the mean flux density at $\lambda = 1010\pm 5$~\AA .
This choice for the continuum bandpass is somewhat arbitrary, but it
avoids the noisy portion of the spectra shortward of $\lambda \approx
1000$~\AA \ and the broad weak bump between the O~VI and C~III $\lambda
1176$ emission lines. Ordered by spin phase, the mean flux density in
this bandpass is $f_{1010} ({10^{-12}~\rm erg~cm^{-2}~s^{-1}~\AA ^{-1}})
=0.181$, 0.163, 0.130, 0.171, and 0.236. Of the spin and binary phases,
it appears that these flux density variations occur on the spin phase,
since as shown in Figure~3 they are reasonably well fitted ($\rm \chi
^2/dof = 6.4/2$ assuming 5\% errors in the flux densities) by $f_{1010}
({10^{-12}~\rm erg~cm^{-2}~s^{-1}~\AA ^{-1}}) = A + B\,\sin 2\pi (\phi
_{67}- \phi_0)$ with $A=0.192\pm 0.005$, $B=0.049 \pm 0.007$, and
$\phi_0=0.743\pm 0.023$; on the binary phase, the fit is significantly
poorer ($\rm \chi ^2/dof = 28.8/2$). The relative FUV continuum pulse
amplitude is therefore $B/A=25\%\pm 4\%$ and the light curve peaks at
$\phi_{67} =-0.01\pm 0.02\approx 0$, consistent with the EUV/soft X-ray
light curve.

Since the {\it ORFEUS\/} bandpass is too narrow to meaningfully constrain
the effective temperature, it is not possible to uniquely determine the
cause of these continuum flux density variations: they could be due to
variations in the effective temperature, variations in the effective size
of the emission region, or some combination of these. Assuming $\Mwd =
0.49\,\Msun $ ($\Rwd = 9.8\times 10^8$~cm), $d=100$~pc, and that the
entire white dwarf surface radiates with a blackbody spectrum, the
effective temperature varies with phase according to $T_{\rm eff}({\rm
kK})=27.2 +1.3\,\sin 2\pi (\phi _{67}-0.743)$. If, as for AM~Her
(\cite{mau98}), we instead assume that we are seeing a 20~kK white dwarf
with 37~kK spot, the apparent projected area of the spot varies with
binary phase according to $f=0.058+0.018\, \sin 2\pi(\phi - 0.743)$. To
demonstrate that such two-temperature blackbody models do a good job of
matching the {\it ORFEUS\/} spectra, we show in Figure~2 a series of
$20+37$ kK blackbody models superposed on the data.

Accompanying the continuum flux variations are variations in the flux and
radial velocity of the emission lines. In what follows, we concentrate
on the emission lines longward of $\lambda = 1000$~\AA\ where the spectra
and hence the line fluxes and positions are not adversely affected by the
high background and consequent low signal-to-noise ratio. Inspection of
Figure~2 reveals that the spectra in the neighborhood of the C~III
$\lambda 1176$ emission line are sufficiently simply to allow fits with a
linear continuum plus a Gaussian (5 free parameters), while the broad and
narrow components of the O~VI $\lambda\lambda 1032$, 1038 doublet require
at a minimum a linear continuum plus four Gaussians. To constrain the
fits of the O~VI line, we constrain the separation of the doublets to
their laboratory separation, and the widths of each component to be the
same (for a total of 10 free parameters).

Figure~4 shows the success we have had with the fits of the O~VI lines,
with both the data and the models binned to a resolution of 0.1~\AA \ and
smoothed with a 5-point triangular filter. Thanks to the high spectral
resolution of the Berkeley spectrograph, the lines are cleanly resolved
into narrow and broad components and it appears that the model produces
reasonable fits of these complex line profiles. The most significant
deviation of the fits relative to the data is in the last spectrum
($\phi_{67}=0.767$--1.152 or $\phi_{98}=0.663$--0.926), where the broad
emission component of the doublet is cut by a pair of slightly
blueshifted ($v<300~\rm km~s^{-1}$) narrow absorption features. A similar
absorption component is present at that same phase in the C~III emission
lines, and it is likely not a coincidence that this absorption is
strongest at the same binary phases where the EUV flux deficit is
strongest ($\phi _{98}\approx 0.85\pm 0.1$). For the present, it is
sufficient to note that the presence of this absorption component does
not appear to significantly affect the fits of the emission lines.

The fitting parameters for the broad and narrow components of the O~VI
emission line have been converted into physical quantities (flux,
velocity, FWHM) and are listed in Table~2. The velocities of the broad
and narrow components of the line were fit with a sinusoidal function
of the form $v=\gamma + K\, \sin 2\pi (\phi-\phi_0)$, whereby it became
apparent that the velocity of the broad component of the line varies
with the spin phase while that of the narrow component varies with the
binary phase. The parameters of these fits are shown in Table~3 and the
data and the best-fit radial velocity curves are shown in Figure~5.
Maximum blueshift of the broad component of the O~VI line occurs at $\phi
_{67}=0.05\pm 0.07\approx 0$, while maximum blueshift of the narrow
component of the line occurs at $\phi _{98}=0.30\pm 0.02\sim 0.25.$ As
shown in Figure~6, these radial velocities anticorrelate nicely with
the flux in the two components of the line. Specifically, the broad-line 
flux varies as $f_{\rm O~VI,~b}\, ({\rm 10^{-12}~erg~cm^{-2}~s^{-1}})=
A+B\,\sin 2\pi (\phi _{67}-\phi_0)$ with $A=3.2\pm 0.2$, $B=1.8\pm
0.2$, and $\phi _0=0.79\pm 0.04$ (hence peaks at $\phi _{67} = 0.04\pm
0.04 \approx 0$), while the narrow-line flux varies as $f_{\rm O~VI,~n}
\, ({\rm 10^{-12}~erg~cm^{-2}~s^{-1}}) = C + D\,\sin 2\pi (\phi _{98}-
\phi_0)$ with $C=0.33\pm 0.03$, $D=0.12\pm 0.03$, and $\phi _0 =0.95\pm
0.07$ (hence peaks at $\phi _{98} = 0.20\pm 0.07 \approx 0.25$).

The behavior of the C~III $\lambda 1176$ emission line is less
straightforward. While the flux in the line clearly correlates with spin
phase according to $f_{\rm C~III}\, ({\rm 10^{-12}~erg~cm^{-2}~s^{-1}}) 
= E + F\,\sin 2\pi (\phi _{67} - \phi_0)$ with $E=1.38\pm 0.09$, $F=0.75
\pm 0.10$, and $\phi _0 = 0.84\pm 0.03$ (hence peaks at $\phi_{67}= 0.09
\pm 0.03\sim 0$), the radial velocity ranges between $\pm 200~\rm
km~s^{-1}$ within the errors and can be fit satisfactorily on either the
spin or binary phases with the parameters shown in Table~3. If the C~III
radial velocity varies with spin phase, it has maximum blueshift at
$\phi _{67}=0.26\pm 0.08$; $\Delta\phi _{67}=0.21\pm 0.11$ {\it after\/}
that of the O~VI broad component, while if the C~III radial velocity
varies with binary phase, it has maximum blueshift at $\phi _{98}=0.14\pm
0.08$; $\Delta\phi _{98}=0.15\pm 0.09$ {\it before\/} that of the O~VI
narrow component. The former alternative is favored by the broad width of
the line and the strong flux variation on the spin phase. However,
because of the long exposures and the relatively poor phase coverage and
because the line is broad and typically rather weak, it is not possible
with the existing data to usefully constrain the phasing of the radial
velocity variations of the C~III $\lambda 1176$ emission line.

\section{IUE Spectroscopy}

As mentioned in the introduction, to date little has been done with the
large number of {\it IUE\/} spectra of EX~Hya in the archive. While a 
full analysis of these UV data is beyond the scope of the present work,
it is nonetheless useful to perform an analysis of a subset of the
existing spectra to extend the bandpass for which we have phase-resolved
spectroscopic information for EX~Hya. The 1995 March {\it HUT\/} spectra
of EX~Hya (\cite{gre97}) of course cover the UV and FUV wavebands
simultaneously, but those observations are limited to some extent by
the limited range of spin ($\phi _{67} = 0.05$--0.50) and binary phases
($\phi _{98}=0.09$--0.40) sampled. Of the 174 {\it IUE\/} spectra in
the archive, 124 are short-wavelength spectra ($\lambda\lambda =
1150$--1950~\AA ) obtained through the large aperture (i.e., are
photometric). Of this subset, there is a continuous set of 45 spectra 
(SWP 55063--55107) with exposure times of 10 minutes obtained by K.\
Mukai over an interval of 1.3 days beginning on 1995 June 23. For the
41 spectra available from the {\it IUE\/} archive, Table~4 lists the
sequence numbers, the HJD of the start of the exposures, and the range of
binary and spin phases assuming the ephemerides of Hellier \& Sproats
(1992).

Unfortunately, even this extensive and continuous set of {\it IUE\/}
spectra suffers from aliasing between the spin and binary periods.
Specifically, the phases of the exposures in this sequence satisfy $\phi
_{67}\approx (0.4-\phi _{98})\pm 0.2$. During the portion of the orbit
unaffected by the dip ($\phi _{98}=0.0$--0.7), there were 14 spectra
obtained during spin maximum ($\phi _{67}=0.8$--1.2), but only 7 spectra
were obtained during spin minimum ($\phi _{67}= 0.3$--0.7); during the
dip ($\phi _{98}=0.75$--0.95), there were 6 spectra obtained during
spin minimum, but none were obtained during spin maximum. To populate
this portion of phase space, we extracted from the archive all (5)
short-wavelength large-aperture spectra satisfying the constraint [$\phi
_{98}=0.75$--0.95, $\phi _{67}= 0.8$--1.2]. The relevant details of these
spectra are included in Table~4.

From this subset of 32 {\it IUE\/} spectra of EX~Hya, we produced the
four mean phase-resolved spectra shown in Figure~7. From brightest to
dimmest, the spectra were obtained during: (1) spin maximum away from the
dip, (2) spin maximum during the dip, (3) spin minimum during the dip, and
(4) spin minimum away from the dip. Like the {\it HUT\/} spectra, these
spectra reveal emission lines of He~II, C~II--IV, N~V, Si~III--IV, and
Al~III superposed on a blue continuum. The most spectacular aspect of
these spectra is the widths of the lines; the FWHM of the C~IV line for
instance is approximately 14~\AA \ or $2700~\rm km~s^{-1}$ compared to
7--10~\AA \ or 1400--$1900~\rm km~s^{-1}$ for other magnetic CVs.

These mean spectra demonstrate the following effects on the UV lines
and continuum as a function of spin and binary phase. First consider the
effect of the dip. During spin maximum, the dip does not significantly
affect the continuum or the He~II line, but the flux in the other lines
decreases by 30\%--40\%, with the red wings of the lines affected
preferentially. During spin minimum, the continuum {\it increases\/} by
roughly 20\% during the dip, but there is little if any effect on the
lines. Next consider variations on the spin phase. Away from the dip, 
the continuum decreases by roughly 40\% going from spin maximum to spin
minimum. The effect on the lines is much more pronounced: the flux in the
N~V and Si~IV lines decreases by roughly 60\%, the flux in the C~IV line
decreases by roughly 80\%, and the He~II line disappears altogether. The
lines also markedly change shape: the N~V and Si~IV lines become less
centrally peaked, while the blue side of the C~IV line is preferentially
suppressed. During the dip, the continuum decreases by roughly 20\% going
from spin maximum to spin minimum, the flux in the C~IV line decreases by
roughly 60\%, again with most of the action on the blue side of the line,
and again the He~II line disappears altogether.

To quantify the variations of the UV lines and continuum as a function
of spin and binary phase, we attempted to fit the flux density of the
individual spectra in the neighborhood of the C~IV line ($\lambda =
1550\pm 50$~\AA ) with a number of analytic functions. Ideally, the C~IV
line in these {\it IUE\/} spectra would be modeled the same way as O~VI
line in the {\it ORFEUS\/} spectra, with a linear continuum plus four
Gaussians, but the {\it IUE\/} spectral resolution is insufficient to
resolve the C~IV line into its doublet components or to separately
distinguish the emission and absorption components. The O~VI narrow
emission and absorption components are relatively weak, so there is some
hope of successfully modeling the C~IV line with a linear continuum plus
one or two Gaussians (with 5 or 8 degrees of freedom, respectively). The
simpler model faithfully measures the continuum flux density, but the
overall fits are poor and the line parameters unreliable because a single
Gaussian is incapable of reproducing the strongly asymmetric shape of the
line during spin minimum. Good fits result if a second Gaussian (either
in emission or absorption) is included in the model, but again the line
parameters are unreliable because they tend to wander in their 
exploration of $\chi ^2$ space. After some experimenting, it was found
that the most robust and reliable line parameters resulted using a model
consisting of a linear continuum plus two Gaussians whose widths were
fixed at 4.0~\AA ; specifically,
$f_\lambda = f_1 + f_2\lambda ^\prime + f_3\, \exp(-[\lambda
^\prime-\lambda_1]^2/2\sigma^2) + f_4\, \exp(-[\lambda
^\prime+\lambda_2]^2/2\sigma^2)$, where $\sigma = 4.0$~\AA , $\lambda
^\prime = \lambda - \lambda _0$, and $\lambda _0 = 1549.48$~\AA , the
optically thick mean of the laboratory wavelengths of the C~IV doublet.

The spin-phase behavior of the resulting flux in the C~IV emission line
is shown in Figure~8 for binary phases during and away from the dip. 
In both cases, the C~IV flux peaks at $\phi _{67}\approx 0$, but the
amplitude and mean level of the oscillation is a strong function of
binary phase. Excluding the anomalously low flux points shown by the
diamonds, away from the dip the C~IV flux varies as $f_{\rm C~IV,~out}\, 
({\rm 10^{-12}~erg~cm^{-2}~s^{-1}})=A + B\sin 2\pi (\phi _{67}-\phi_0)$,
where $A=5.6\pm 0.2$, $B=4.0\pm 0.3$, and $\phi_0=0.76\pm 0.1$ (hence
peaks at $\phi_{67}=0.01\pm 0.01\approx 0$), whereas during the dip the
flux varies as $f_{\rm C~IV,~in}\, ({\rm 10^{-12}~erg~cm^{-2}~s^{-1}}) =
C + D\, \sin 2\pi (\phi _{67}-\phi_0)$, where $C=3.9\pm 0.2$, $D=1.6\pm 
0.2$, and $\phi_0=0.76\pm 0.04$ (hence peaks at $\phi _{67}=0.01\pm 0.04
\approx 0$). The spin-phase variation of the C~IV continuum flux density
(specially, the model flux density at $\lambda = 1549.48$~\AA ) is shown
in Figure~9. At least away from the dip, there is a tendency for the
continuum flux density to be higher near $\phi _{67}=0$ and lower near
$\phi _{67}=0.5$, but there is considerable scatter in the data at any
spin phase.

The C~IV line widths and radial velocities also follow from this
parameterization of the spectra, although indirectly: the radial 
velocity is $v=c\, (\lambda_ 1-\lambda _2)/\lambda _0$ and the line width
(strictly, the separation of the two Gaussians) is $w=c\, (\lambda _1
+\lambda _2)/\lambda _0$. The most obvious variation is that of $w$, 
which varies with spin phase as $w\, ({\rm 10^3~km~s^{-1}}) = \gamma+K\,
\sin 2\pi (\phi_{67}-\phi _0)$, with $\gamma =1.9\pm 0.03$, $K=0.34\pm
0.04$, and $\phi _0=0.20\pm 0.02$ (hence peaks at $\phi _{67}=0.45\pm
0.02\sim 0.5$), but this variation is caused by the model's fitting of
the single-peaked line profiles (i.e., small Gaussian separations) during
spin maximum and the double-peaked line profiles (i.e., large Gaussian
separations) during spin minimum and does not translate into a variation
in the net width of the line: the FWZI of the line is instead reasonably
constant (with only a few exceptions, within 10\%) at 30~\AA \ or
$5700~\rm km~s^{-1}$. Like the C~III $\lambda 1176$ line (but unlike the
O~VI line) in the {\it ORFEUS\/} spectra, there is no radial velocity
variation of the C~IV line apparent on the white dwarf spin phase.
However, as seen already in Figure~7, there is a tendency for the line to
shift toward the blue during the dip (although the scatter in the
individual measurements is large and the velocity difference is formally
consistent with zero [$v=-680\pm 710~\rm km~s^{-1}$ during the dip
compared to $v=180\pm 1600~\rm km~s^{-1}$ away from the dip, where the
errors are the square root of the sample variance relative to the
weighted mean]). If the C~IV radial velocity were as large as that of the
O~VI broad component, its amplitude would be roughly 1.7~\AA , which is
comparable to the width of the wavelength bins in the {\it IUE \/}
spectra. Evidently, the complexity of the C~IV line combined with the low
spectral resolution and modest signal-to-noise ratio of the {\it IUE\/}
spectra preclude centroiding the line to determine, or place useful
limits on, its radial velocity.

\section{Discussion}

Before wading into details, it is useful to compare the mean {\it
ORFEUS\/} spectrum of EX~Hya with that of AM~Her assembled from the six
spectra obtained days earlier with the same instrument (\cite{mau98},
hereafter MR98). These mean FUV spectra are shown in Figure~10, where
the spectrum of AM~Her has been multiplied by $(75/100)^2$ to account for
the relative distance to the two sources. Given that EX~Hya (an IP with
a truncated accretion disk) and AM~Her (a polar without a disk) are
physically such different sources, it is amazing that their FUV spectra
are so similar. First, the {\it level\/} of the FUV continua are nearly
identical. Second, the {\it shapes\/} of the FUV continua are nearly
indistinguishable, even so far as (1) the absence of Lyman absorption
lines and (2) the presence of the broad weak bump between the O~VI and
C~III $\lambda 1176$ emission lines. Third, both sources have C~III,
N~III, and O~VI emission lines with comparable widths and intensities;
indeed, the intrinsic flux in the C~III $\lambda 977$ and N~III $\lambda
991$ emission lines are nearly identical. Fourth, both sources show broad
and narrow component structure in the O~VI emission line. With the
exception of the absence of the He~II $\lambda 1085$ emission line in the
spectrum of EX~Hya, the {\it differences\/} between these spectra are
in the details. First, the broad (narrow) component of the O~VI line of
EX~Hya is stronger (weaker) than that of AM~Her. Second, the C~III
$\lambda 1176$ emission line of EX~Hya is brighter and broader than that
of AM~Her.

Next consider the constraints imposed on the location of the continuum
and emission-line regions by the phase-resolved {\it ORFEUS\/} spectra.
First consider the broad-line region. MR98 identified the accretion
funnel as the source of the broad component of the O~VI emission line
in their {\it ORFEUS\/} spectra of AM~Her. Consistent with simple
expectations, in AM~Her spin maximum of the FUV continuum and X-ray
light curves occurs when the upper pole points {\it toward\/} the
observer---when the {\it redshift\/} of the O~VI broad component is the
highest. Similarly, we identify the accretion funnel as the source of the
O~VI broad emission component in our {\it ORFEUS\/} spectra of EX~Hya,
but, consistent with the accretion curtain model, spin maximum of the FUV
continuum and X-ray light curves occurs when the upper pole points {\it
away\/} the observer---when the {\it blueshift\/} of the O~VI broad
component is the highest. This geometry implicates the accretion funnel
itself as the source of the FUV continuum flux, not a separate
intermediate-temperature spot on the surface of the white dwarf. If
this is the case for both sources, it solves the problem of the absence
of Lyman absorption lines in their FUV spectra. Next consider the
narrow-line region. MR98 identified the irradiated face of the secondary
as the source of the narrow component of the O~VI emission line in their
{\it ORFEUS\/} spectra of AM~Her. The secondary cannot be the site of the
narrow-line region in EX~Hya, however, because maximum blueshift of the
O~VI narrow emission component occurs when the {\it white dwarf\/}, not
the {\it secondary\/}, is moving toward the observer. Indeed, the radial
velocity solution of the O~VI narrow emission component ($K=85\pm 9~\rm
km~s^{-1}$, $\phi_0= 0.54\pm 0.02$; Table~3) is consistent with that of
the Balmer line wings in the optical ($K=69\pm 9~\rm km~s^{-1}$, $\phi_0
=0.53\pm 0.03$; \cite{hel87}), so we identify the white dwarf itself as
the source of the O~VI narrow emission component. With a $\rm FWHM\approx
200~\rm km~s^{-1}$, the O~VI narrow emission component may ultimately
prove to be better than the Balmer lines (for which $\rm FWHM\sim
2000~\rm km~s^{-1}$) for determining the radial velocity of the white
dwarf in EX~Hya.

Delving further into details, it is possible to combine information
from the {\it ORFEUS\/} and {\it IUE\/} spectra of EX~Hya to constrain
the physical condition of the FUV and UV line-emitting plasma. First
consider the broad-line region. In dense gas illuminated by hard X-rays
(model 4 of \cite{kal82}), O~VI exists over a range of ionization
parameters $\xi\equiv L/nr^2\approx 40$--70 and temperatures $T\approx
40$--130 kK. With $L\approx 2\times 10^{32}~\rm erg~s^{-1}$
(\cite{all98}), $n\gax 2\times 10^{11}~\rm cm^{-3}$ for $r\le 5\times
10^9~{\rm cm}\approx 5\, \Rwd$. For the inferred range of ionization
parameters, the dominant ionization stages of He, C, N, and Si are He
II--III, C VI--VII, N VI--VII, and Si VI--XI, respectively, so if the
observed lower ionization species dominate they must be produced in gas
which is denser and/or lies further from the source of the ionizing 
flux. The observed line ratios may be affected by finite optical depths 
in the resonance lines, but modulo this effect the mean C~III $\lambda 
977/\lambda 1176$ line ratio of $\sim 1.4$ requires $n\gax 10^{11}~\rm
cm^{-3}$ and $T\gax 80$~kK (\cite{kee92}; \cite{kee97}) and the mean
Si~III $\lambda 1300/\lambda 1890$ line ratio of $\gax 10$ requires
$n\gax 10^{12}~\rm cm^{-3}$ and $T\gax 60$~kK (\cite{nus86}). The He~II
$\lambda 1085/\lambda 1640$ line ratio presents a puzzle. During spin
maximum, the strength of the $\lambda 1640$ line in the {\it IUE\/}
spectra is $f_{1640}\approx 1.5\times 10^{-12}~\rm erg~cm^{-2}~s^{-1}$
(consistent with the {\it HUT\/} measurement), and the strength of the
$\lambda 1085$ line in the {\it ORFEUS\/} spectra is $f_{1085}\lax
0.2\times 10^{-12}~\rm erg~cm^{-2}~s^{-1}$ (a factor of $\gax 2$ less 
than the uncertain {\it HUT\/} estimate), so the He~II $\lambda 1085/
\lambda 1640$ line ratio is $\lax 0.13$. For case B recombination, this
ratio is $>0.13$ for the full range of densities ($n =
10^{2}$--$10^{14}~\rm cm^{-3}$) and temperatures ($T=10$--100 kK)
tabulated by Storey \& Hummer (1995), and is $>0.17$ for $n>10^{10}~\rm
cm^{-3}$ and $T<100$ kK. Variability could explain this discrepancy, but
if it does not we must appeal to some process which preferentially
destroys $\lambda 1085$ line photons and/or enhances $\lambda 1640$ line
photons. If the population of the $n=2$ level is high enough to render
the $\lambda 1085$ transition optically thick, there is a branching 
ratio of order one half to convert $\lambda 1085$ photons into a
combination of He~II Balmer ($\lambda 1216$, $\lambda 1640$), Paschen
($\lambda 4686$), and Brackett photons; this process would roughly double
the flux of $\lambda 1640$ photons and thereby resolve the discrepancy.
Simultaneously, it is possible for the He~II Balmer $\beta $ transition
to be pumped by H~I Lyman $\alpha $ photons, which generates $\lambda
1640$ and $\lambda 4686$ line photons when the ion decays. The strength
of the $\lambda 4686$ line in the phase-averaged optical spectrum of
Hellier et al.\ (1987) is uncertain because the absolute flux calibration
is uncertain, but the measured value is $f_{4686}\sim 0.2\times
10^{-12}~\rm erg~cm^{-2}~s^{-1}$, so the He~II $\lambda 4686/\lambda 
1640$ line ratio is $\sim 0.13$. This ratio is consistent with the case B
line ratios of Storey \& Hummer, but it is unfortunately not diagnostic,
given the uncertain calibration of the optical spectrum. Nonetheless, the
available evidence points to the $\lambda 1085$ line flux being lower
than expected, but since there are many ways for this to come about, in
the absence of a detailed model it does not constrain the plasma
conditions.

Next consider the narrow-line region, or more specifically why the
irradiated face of the secondary of AM~Her produces a narrow O~VI 
emission line, while that of EX~Hya does not. Shielding of the secondary
by the accretion disk may play a role, but we argue that the fundamental
reason is that the ionization parameter is simply too low. Although the
luminosity of the hard component of the X-ray spectra of both AM~Her
and EX~Hya is $L\approx 2\times 10^{32}~\rm erg~s^{-1}$ (\cite{ish97};
\cite{all98}), AM~Her also has a soft component in its X-ray spectrum
with $L\approx 2\times 10^{33}~\rm erg~s^{-1}$ (\cite{pae96}). The ten
times lower net ionizing luminosity of EX~Hya is exacerbated by the
lower efficiency of photoionization by hard X-rays, but is ameliorated
by the factor of two smaller distance from the white dwarf to the face
of the secondary. Specifically, whereas the ionization parameter of
the irradiated face of the secondary of AM~Her is $\xi\approx 2\times
10^{33}~{\rm erg~s^{-1}}/2\times 10^{10}~{\rm cm^{-3}} /(5.3\times
10^{10}~{\rm cm})^2 = 36$, for which O~VI dominates for dense gas
illuminated by a mixture of hard and soft X-rays (model 5 of
\cite{kal82}), that of EX~Hya is $\xi\approx 2\times 10^{32}~{\rm
erg~s^{-1}}/2\times 10^{10}~{\rm cm^{-3}}/(2.7\times 10^{10}~{\rm cm})^2
= 14$, for which O~I dominates for dense gas illuminated by hard X-rays
alone (model 4 of \cite{kal82}). For O~VI to dominate in EX~Hya, the
plasma must lie closer to the source of the ionizing flux. To satisfy the 
phasing of the radial velocity of the O~VI narrow emission component, the
narrow-line region must be closer to the white dwarf than the center of
mass of the binary ($r<6.7\times 10^9~\rm cm$), hence $n\gax 10^{11}~\rm
cm^{-3}$, consistent with the density derived above for the broad-line
region. Based on the ratio $R$ of the O~VI line intensities shown in
Table~2, the narrow-line region is optically thick ($R=0.92\pm 0.29$),
while the broad-line region is more likely optically thin ($R=1.8\pm
1.0$).

\section{Summary}

Using {\it EUVE\/} photometry and {\it ORFEUS\/} and {\it IUE\/}
spectroscopy, we have presented a detailed picture of the behavior of
EX~Hya in the vacuum ultraviolet. Consistent with its behavior in the
optical, and hence consistent with the accretion curtain model of EX~Hya,
we find that the FUV and UV continuum flux densities, the FUV and UV 
broad emission line fluxes, and the radial velocity of the O~VI broad
emission component all vary on the spin phase of the white dwarf, with
the maximum of the FUV and UV continuum and broad emission line flux
light curves coincident with maximum {\it blueshift\/} of the broad O~VI
emission component. On the binary phase, we find that the strong eclipse
of the EUV flux by the bulge on the edge of the accretion disk is
accompanied by narrow and relatively weak absorption components of the
FUV emission lines and 30\%--40\% eclipses of all the UV emission lines
except He~II $\lambda 1640$, while the UV continuum is largely unaffected.
Furthermore, both the flux and radial velocity of the O~VI narrow
emission component vary with binary phase. From the relative phasing of
the FUV and UV continuum light curves and the FUV emission-line radial
velocities, we identify the accretion funnel as the source of the FUV and
UV continuum and the O~VI broad emission component, and the white dwarf
as the source of the O~VI narrow emission component. The irradiated face
of the secondary of EX~Hya does not produce the narrow O~VI emission
component observed in {\it ORFEUS\/} spectra of AM~Her because the
ionization parameter (the X-ray luminosity) is too low. Various lines
of evidence imply that the density of both the broad- and narrow-line
regions is $n\gax10^{11}~\rm cm^{-3}$, but the O~VI line ratios
imply that the narrow-line region is optically thick while the broad-line
region is more likely optically thin. As in AM~Her, it is likely that the
velocity shear in the broad-line region allows O~VI photons to escape,
rendering the gas effectively optically thin.

\acknowledgments

We thank all those involved with making the {\it ORFEUS-SPAS II\/}
mission a success: the members of the satellite and instrument teams
at Institute for Astronomy and Astrophysics, University of T\"ubingen;
Space Science Laboratory, University of California, Berkeley; and
Landessternwarte Heidelberg-K\"onigstuhl; the flight operations team;
and the crew of STS-80. Special thanks (and apologies) are due to 
K.~Mukai for acquiring the extensive set of {\it IUE\/} spectra used
herein. F.~Keenan is warmly thanked for supplying C~III level population
and line intensity data. Conversations and correspondence with
B.~Greeley, C.~Hellier, K.~Long, J.~Raymond, and M.~Sirk are gratefully
acknowledged. This work was performed under the auspices of the
U.S.~Department of Energy by Lawrence Livermore National Laboratory under
contract No.~W-7405-Eng-48.

%The anonymous referee is acknowledged for helpful comments.

\clearpage % force page break

% Tables
%---------------------------------------------------------
%redefine ! to use in the tables as a spacer: 

\newdimen\digitwidth
\setbox0=\hbox{\rm0}
\digitwidth=\wd0
\catcode`!=\active
\def!{\kern\digitwidth}

% Table 1
%---------------------------------------------------------

\begin{table*}
\begin{center}
\begin{tabular}{cccc}
\multicolumn{4}{c}{TABLE 1} \\
\multicolumn{4}{c}{Journal of {\it ORFEUS\/} Observations} \\
\tableline \tableline
Start Date& Exposure& \\
($\rm HJD-2450000$) & (s)& $\phi_{98}$& $\phi_{67}$\\
\tableline
419.37867&  1717&  !0.497--!0.788&  !0.341--!0.767\\
419.56674&  2018&  !3.253--!3.595&  !4.381--!4.882\\
419.68857&  2328&  !5.038--!5.433&  !6.998--!7.577\\
419.75218&  1342&  !5.971--!6.208&  !8.365--!8.713\\
421.02764&  1549&  24.663--24.926&  35.767--36.152\\
\tableline
\end{tabular}
\end{center}
\end{table*}
\clearpage % force page break

% Table 2
%---------------------------------------------------------

% Table 2 must be \LaTeX ed separately, printed sideways, and inserted
% here; increment page counter so that this can be accomplished.......

\addtocounter{page}{1}

% Table 3
%---------------------------------------------------------

\begin{table*}
\begin{center}
\begin{tabular}{lcccc}
\multicolumn{5}{c}{TABLE 3} \\
\multicolumn{5}{c}{Radial Velocities} \\
\tableline \tableline
    &          & $\gamma $&         $K$&                        \\
Line& Component& $\rm (km~s^{-1})$& $\rm (km~s^{-1})$& $\phi _0$\\
\tableline
O VI  $\lambda\lambda 1032$, 1038& N\tablenotemark{a}& $+45\pm !6$&  $!85\pm !9$&  $0.54\pm 0.02$\\
O VI  $\lambda\lambda 1032$, 1038& B\tablenotemark{b}& $-79\pm 59$&  $332\pm 65$&  $0.30\pm 0.07$\\
C III $\lambda        1176$      & B\tablenotemark{a}& $-40\pm 40$&  $118\pm 56$&  $0.39\pm 0.08$\\
C III $\lambda        1176$      & B\tablenotemark{b}& $-40\pm 39$&  $113\pm 47$&  $0.51\pm 0.08$\\
\tableline
\end{tabular}
\tablenotetext{a}{$v=\gamma + K\, \sin\, 2\pi (\phi_{98} - \phi _0)$.}
\tablenotetext{b}{$v=\gamma + K\, \sin\, 2\pi (\phi_{67} - \phi _0)$.}
\end{center}
\end{table*}

\clearpage % force page break

% Table 4
%---------------------------------------------------------

\begin{table*}
\begin{center}
\begin{tabular}{cccc}
\multicolumn{4}{c}{TABLE 4} \\
\multicolumn{4}{c}{Journal of {\it IUE\/} Observations\tablenotemark{a}}\\
\tableline \tableline
Sequence& Start Date         &             &            \\
Number  & ($\rm HJD-2440000$)&  $\phi_{98}$& $\phi_{67}$\\
\tableline
SWP 17598& 5187.15547&  0.736--!0.889&  0.999--!0.223\\
SWP 23199& 5859.18716&  0.685--!0.889&  0.937--!0.235\\
SWP 26547& 6282.38720&  0.887--!0.009&  0.976--!0.155\\
SWP 28858& 6649.95322&  0.744--!0.151&  0.775--!0.372\\
SWP 47643& 9118.84774&  0.587--!0.009&  0.630--!0.249\\
SWP 55063& 9892.07765&  0.645--!0.747&  0.760--!0.909\\
SWP 55064& 9892.10641&  1.067--!1.168&  1.378--!1.527\\
SWP 55065& 9892.13367&  1.466--!1.568&  1.964--!2.113\\
SWP 55066& 9892.16015&  1.854--!1.956&  2.532--!2.681\\
SWP 55067& 9892.18676&  2.244--!2.346&  3.104--!3.253\\
SWP 55068& 9892.21373&  2.639--!2.741&  3.683--!3.832\\
SWP 55069& 9892.23968&  3.020--!3.121&  4.241--!4.390\\
SWP 55070& 9892.27026&  3.468--!3.570&  4.898--!5.047\\
SWP 55071& 9892.29733&  3.865--!3.966&  5.480--!5.629\\
SWP 55073& 9892.35149&  4.658--!4.760&  6.643--!6.792\\
SWP 55074& 9892.39783&  5.338--!5.439&  7.639--!7.788\\
SWP 55075& 9892.42914&  5.796--!5.898&  8.311--!8.460\\
SWP 55076& 9892.46114&  6.265--!6.367&  8.999--!9.148\\
SWP 55077& 9892.49171&  6.713--!6.815&  9.656--!9.805\\
SWP 55078& 9892.52777&  7.242--!7.344& 10.430--10.579\\
SWP 55079& 9892.55893&  7.699--!7.800& 11.100--11.249\\
SWP 55080& 9892.59520&  8.230--!8.332& 11.879--12.028\\
SWP 55081& 9892.62656&  8.690--!8.791& 12.553--12.702\\
%\tableline
\end{tabular}
\end{center}
\end{table*}

\begin{table*}
\begin{center}
\begin{tabular}{cccc}
%\multicolumn{4}{c}{TABLE 4 ({\it continued})} \\
%\multicolumn{4}{c}{Journal of {\it IUE\/} Observations} \\
%\tableline \tableline
%Sequence& Start Date         &             &            \\
%Number  & ($\rm HJD-2440000$)&  $\phi_{98}$& $\phi_{67}$\\
%\tableline
SWP 55082& 9892.65456&  9.100--!9.202& 13.154--13.303\\
SWP 55083& 9892.68347&  9.524--!9.625& 13.775--13.924\\
SWP 55084& 9892.71699& 10.015--10.117& 14.495--14.644\\
SWP 55086& 9892.77016& 10.794--10.896& 15.638--15.787\\
SWP 55087& 9892.79685& 11.185--11.287& 16.211--16.360\\
SWP 55088& 9892.82308& 11.570--11.671& 16.775--16.924\\
SWP 55089& 9892.84972& 11.960--12.062& 17.347--17.496\\
SWP 55091& 9892.90274& 12.737--12.839& 18.486--18.635\\
SWP 55092& 9892.93000& 13.137--13.238& 19.072--19.221\\
SWP 55093& 9892.95751& 13.540--13.642& 19.663--19.812\\
SWP 55094& 9892.98448& 13.935--14.037& 20.242--20.391\\
SWP 55095& 9893.01235& 14.344--14.445& 20.841--20.990\\
SWP 55097& 9893.06511& 15.117--15.219& 21.974--22.124\\
SWP 55098& 9893.09177& 15.508--15.609& 22.547--22.696\\
SWP 55099& 9893.12323& 15.969--16.070& 23.223--23.372\\
SWP 55100& 9893.15120& 16.379--16.480& 23.824--23.973\\
SWP 55101& 9893.17884& 16.784--16.885& 24.418--24.567\\
SWP 55102& 9893.20700& 17.196--17.298& 25.023--25.172\\
SWP 55103& 9893.23465& 17.602--17.703& 25.617--25.766\\
SWP 55104& 9893.26278& 18.014--18.116& 26.221--26.370\\
SWP 55105& 9893.28999& 18.413--18.514& 26.806--26.955\\
SWP 55106& 9893.31825& 18.827--18.928& 27.413--27.562\\
SWP 55107& 9893.35007& 19.293--19.395& 28.097--28.246\\
\tableline
\end{tabular}
\tablenotetext{a}{Exposures are as follows. SWP 17598: 900~s,
SWP 23199: 1200~s, SWP 26547: 721~s, SWP 28858: 2400~s, SWP 
47643: 2489~s, SWP 55063--55107: 600~s.}
\end{center}
\end{table*}

\clearpage % force page break

% References
%---------------------------------------------------------

\clearpage % force page break

% Figure captions
%---------------------------------------------------------

\figcaption%[figure1.ps]
{Spin- ({\it upper panel}) and binary-phase ({\it middle and lower
panels}) {\it EUVE\/} deep survey light curves of EX~Hya. One hundred
phase bins are used in the upper two panels and 589 (10 second bins) 
are used in the lower panel. In the upper two panels a typical error bar 
($\pm 0.012~\rm counts~s^{-1}$) is indicated by the cross. The numbered
horizontal  lines indicate the relative phases of the {\it ORFEUS\/}
spectra. For reference, $0.2~\rm counts~s^{-1}\approx 2 \times
10^{-11}~\rm erg~cm^{-2}~s^{-1}$.
\label{fig1}}

\figcaption%[figure2.ps]
{{\it ORFEUS\/} spectra of EX~Hya ordered by white dwarf spin phase. Each
successive  spectrum is offset by 0.75 flux density units. Two-component
($20+37$ kK) blackbody models are shown by the light-colored nearly
straight curves.
\label{fig2}}

\figcaption%[figure3.ps]
{Mean flux density at $\lambda = 1010\pm 5$~\AA \ in units of $\rm
10^{-12}~erg~cm^{-2}~s^{-1}~\AA ^{-1}$ as a function of white dwarf spin
phase.
\label{fig3}}

\figcaption%[figure4.ps]
{Regions of the {\it ORFEUS\/} spectra containing the O~VI doublet
showing Gaussian fits to the broad and narrow components. Panels are
ordered by relative white dwarf spin phase.
\label{fig4}}

\figcaption%[figure5.ps]
{{\it Upper panel\/}: Radial velocity of the broad component of the O~VI
doublet as a function of white dwarf spin phase. {\it Lower panel\/}:
Radial velocity of the narrow component of the O~VI doublet as a function
of binary phase.
\label{fig5}}

\figcaption%[figure6.ps]
{{\it Upper panel\/}: Flux of the broad component of the O~VI doublet
as a function of white dwarf spin phase. {\it Lower panel\/}: Flux of
the narrow component of the O~VI doublet as a function of binary phase.
\label{fig6}}

\figcaption%[figure7.ps]
{Mean phase-resolved {\it IUE\/} spectra of EX~Hya for the following
spin/binary phases: ({\it a\/}) maximum/non-dip, ({\it b\/}) maximum/dip,
({\it c\/}) minimum/dip, and ({\it d\/}) minimum/non-dip. Reseaux are
marked by crosses ($\times$).
\label{fig7}}

\figcaption%[figure8.ps]
{Flux of the C~IV doublet as a function of white dwarf spin phase away
({\it upper panel}) and during ({\it lower panel}) the dip in the orbital
light curve.
\label{fig8}}

\figcaption%[figure9.ps]
{Flux density at $\lambda = 1549.48$~\AA \ as a function of white dwarf
spin phase away ({\it upper panel}) and during ({\it lower panel}) the
dip in the orbital light curve.
\label{fig9}}

\figcaption%[figure10.ps]
{Mean {\it ORFEUS\/} spectra of EX~Hya ({\it thick histogram\/}) and
AM~Her [multiplied by $(75/100)^2$ to account for the relative distance 
to the two sources] ({\it light-colored histogram\/}).
\label{fig10}}

%---------------------------------------------------------
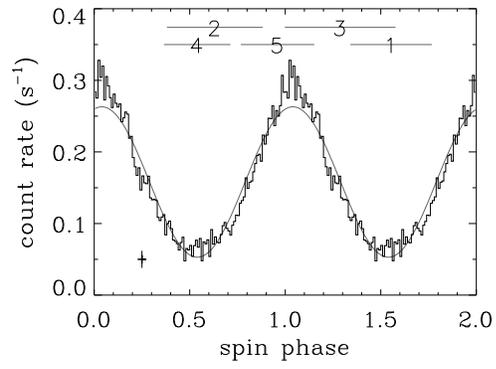
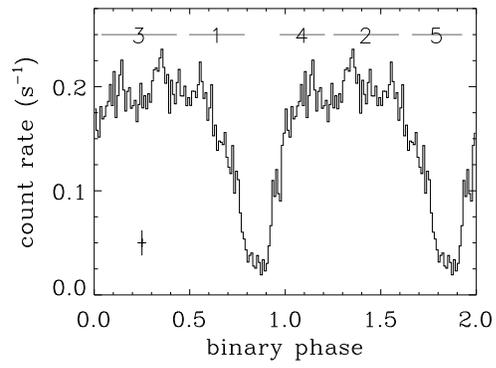
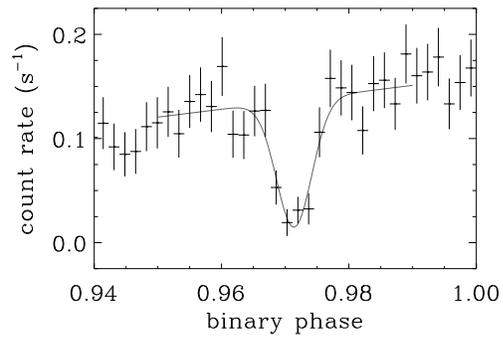
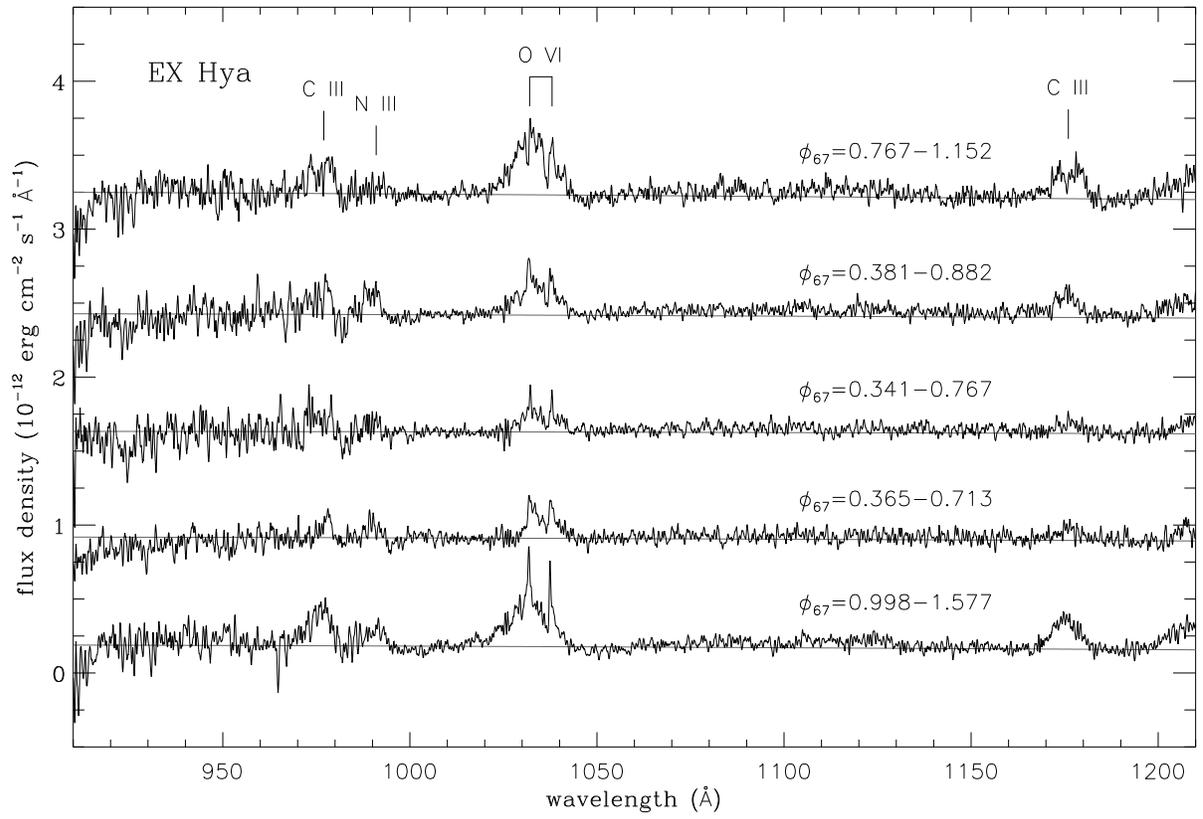
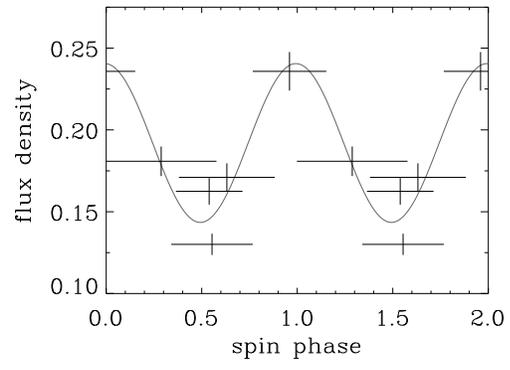
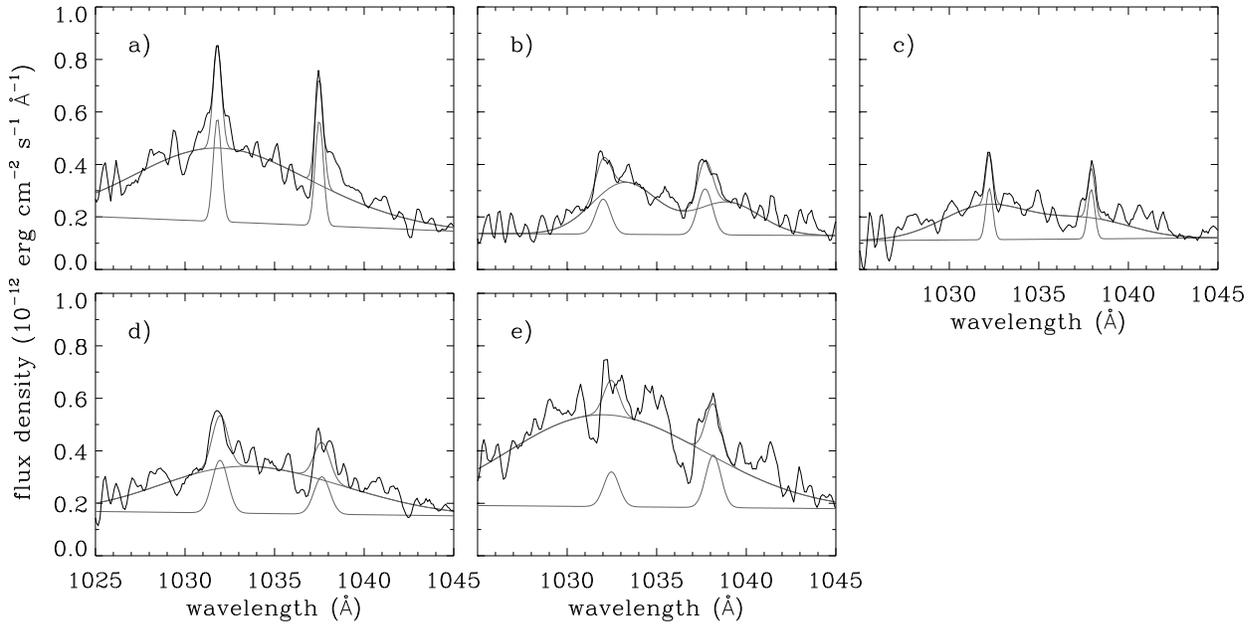
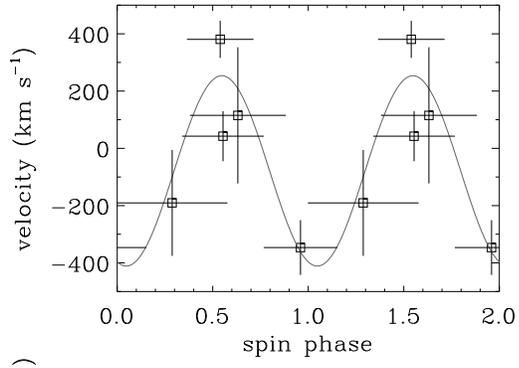
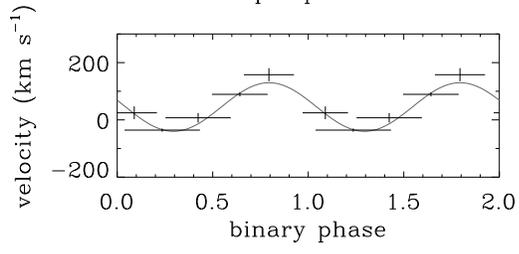
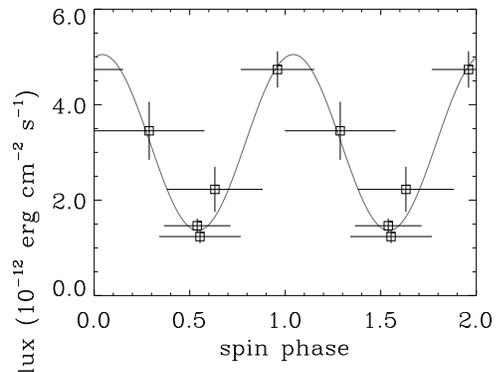
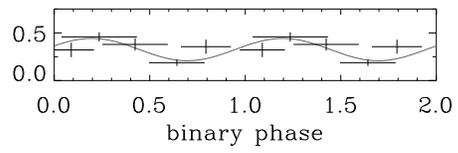
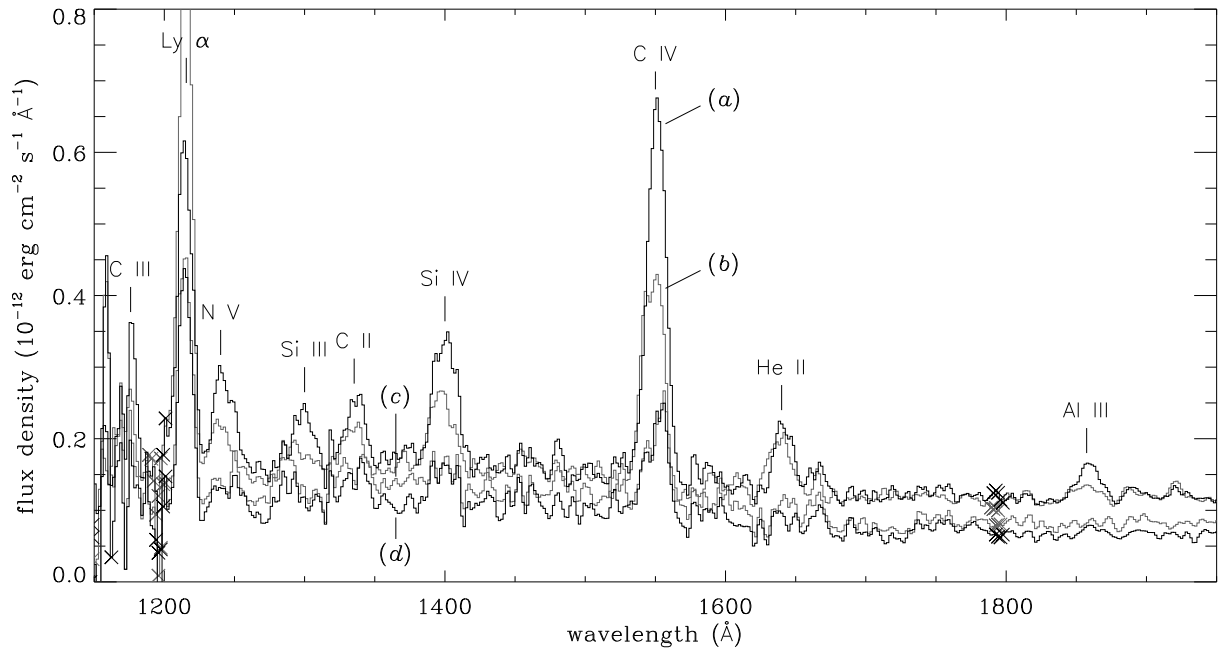
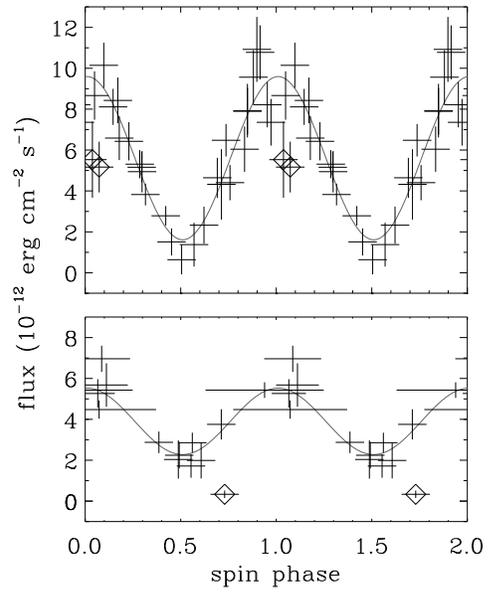
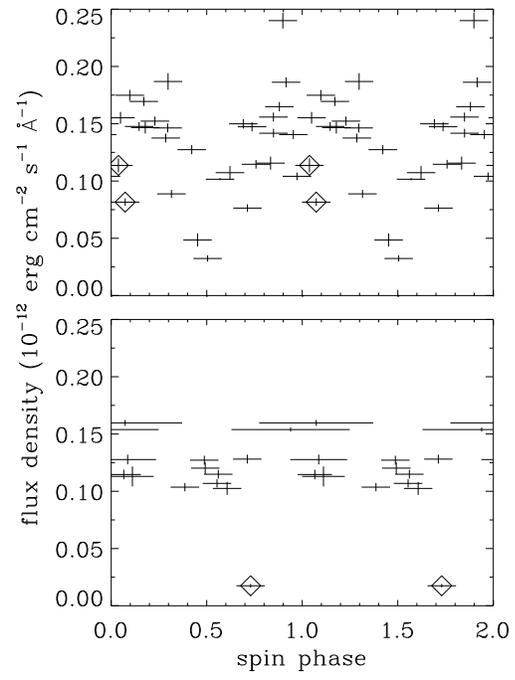
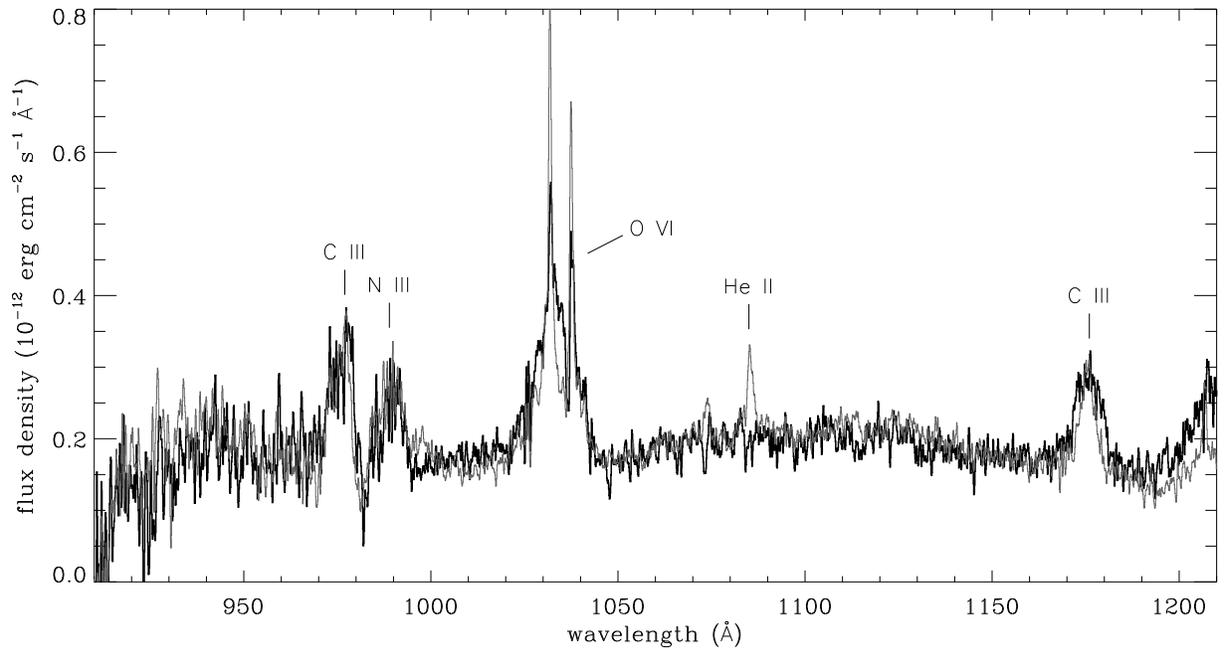
\end{document}